\RequirePackage{fixltx2e}
\documentclass[aps,prb,twocolumn,floatfix]{revtex4-1}

\usepackage{dcolumn,amsfonts,amsmath,amssymb}
\usepackage{bm}
\usepackage{lipsum}
\usepackage{graphicx}
\usepackage{epsfig}
\usepackage{float}
\usepackage[export]{adjustbox}

\usepackage{graphicx}
\begin{document}
\title{Theoretical investigation of charge transport in germanium doped phosphorene nanoribons using DFT + NEGF}
\author{Maryam Azizi}
\author{Badie Ghavami}
\affiliation{School of Nanoscience, Institute for Research in fundamental Sciences, P.O.Box: 19395-5531, Tehran, Iran}
\date{\today}

\begin{abstract}
\label{abstract}
New two diemensional structures nanoribbon including phosphorus and germanium atoms are introduced 
for the nanoelectronic applications. 
Under various 
bias voltages, the electronic transport in the systems have been studied within the noneqilibrium Green's function formalism. 
The $I-V$ characteristics have been extracted. 
DOS and $T(E,V_{bias})$ have been investigated and show that the charge transport occurs 
when the bias voltage reaches about 1 \textit{V}. 
The calculated MPSH shows that the spatial
distribution of orbital levels has been affected by the electrodes. 
The studied structures have a bandgap of about 0.7 \textit{eV} which absorbs light in 
the visible range and thus could be an interesting contender for solar cells applications.
\end{abstract}

\maketitle

\section{Introduction}
\label{introduction}
Two dimensional (2D) graphene-like materials, have recently attracted considerable attention due to 
their potential applications in nano- and optoelectronics\cite{davila2016few,bonaccorso2010graphene,schwierz2010graphene,nicolosi2013liquid}. 
In particular, their unique size dependent properties allowed for the exploration of 
a large number of novel phenomena at nanoscale\cite{wang2015black,saroka2017electro,halliouglu2015formation,maity2016structural}.
Among 2D materials, the ones with sizable bandgap are used in field effect transistor (FET) 
devices. In the recent years, phosphorene\cite{guo2017modulation,lv2016scaling,carvalho2014phosphorene} (monolayer of black phosphorus) has attracted great attention 
due to its reasonable mobility and bandgap which makes it an attractive material for electronic applications. 
The effects of native defects\cite{wang2015native},
vacancies, and adatoms\cite{srivastava2015tuning} have been investigated for this material. 
More over, tunability of electronic properties, for example
due to strain\cite{peng2014strain,elahi2015modulation,jiang2015analytic}, 
paves the way to interesting applications of the material. 
In addition to phosphorene sheets,
phosphorene nanoribbons (PNR) have also been studied 
theoretically\cite{jiang2015analytic,zhang2014phosphorene,maity2016structural}.

Another promissing material with a tunable bandgap is 
germanene\cite{ni2011tunable,drummond2012electrically} (monolayer of germanium atoms)

Although both phoshorene and germanene nanoribbons show 
certain advantages in comparison to graphene and 
2D dichalcogenides such as their application in energy conversion/storage and 
high performance field effect transistors (FET), there have been few transport studies compared to 
graphene and $MoS_2$\cite{liu2015semiconducting, das2014ambipolar, ghavami2015varistor,kamalakar2015low}. Moreover, to our 
knowledge, the only recent work on the combination of these two elements\cite{jing2017gep3} 
considers the n layer of $GeP_3$ and focuses on the 
low indirect bandgaps and high carrier mobilities.
Since monolayers of germanium atoms show semimetallic properties, combining this 
element with phosphorus which is known as a wide bandgap material may result in a moderate bandgap material.
Hence, in this letter, we for the first time introduce two nanoribbon structures composed of both germanium and 
phosphorus atomes. These new nanoribbons have a bandgap about 0.7 \textit{eV} which absorbs light 
in the visible range and could be suitable for solar cells applications. 
Non-equilibrium Green's function method based on density functional theory (DFT) has been used to study charge and quantum transport properties. Applying bias voltage, Density of States (DOS) and transmission spectrum as a function of energy have been investigated. 


The paper is organized as follows. In Sec.\ref{sec:Computational_Details}, we 
first introduce the system and explain the method we use to derive current-voltage 
characteristics, density of states (DOS) and transmission spectrum. 
Then in Sec.\ref{sec:results}, we present our numerical results. In particular, the role 
of bias voltage on the transmission spectrum has been shown.
Finally, we conclude and summarize the main
achivement in Sec.\ref{sec:concl}.

\section{Model}

\label{sec:Computational_Details}
\subsection{Structure}
In this work we study a nanoribbon structures composed of 
germanium and phosphorus atoms, with a bandgap between pure germanene 
and pure phosphorene nanoribbon, depicted in Fig.\ref{structure}. \\
In this model phosphorus atoms in the both zigzag edges of the phoshorene nanoribbon, Fig.\ref{structure}a,
were replaced by germanium atoms to construct Fig.\ref{structure}c. To 
go further with the investigation of the effect of germanium atoms, the second chain of phosphorus atomes was replaced 
by germanium ones Fig.\ref{structure}c.
We assumed that each edge atom was passivated with enough H to remove dangling bonds.

This replacement decreases the bandgap of the pure phosphorene nanoribbon and makes it more suitable for 
the certain electronic applications discussed in Sec.\ref{sec:results}.\\

\begin{widetext}
\begin{center}
\begin{figure}[h]
	\includegraphics[scale=0.35,center]{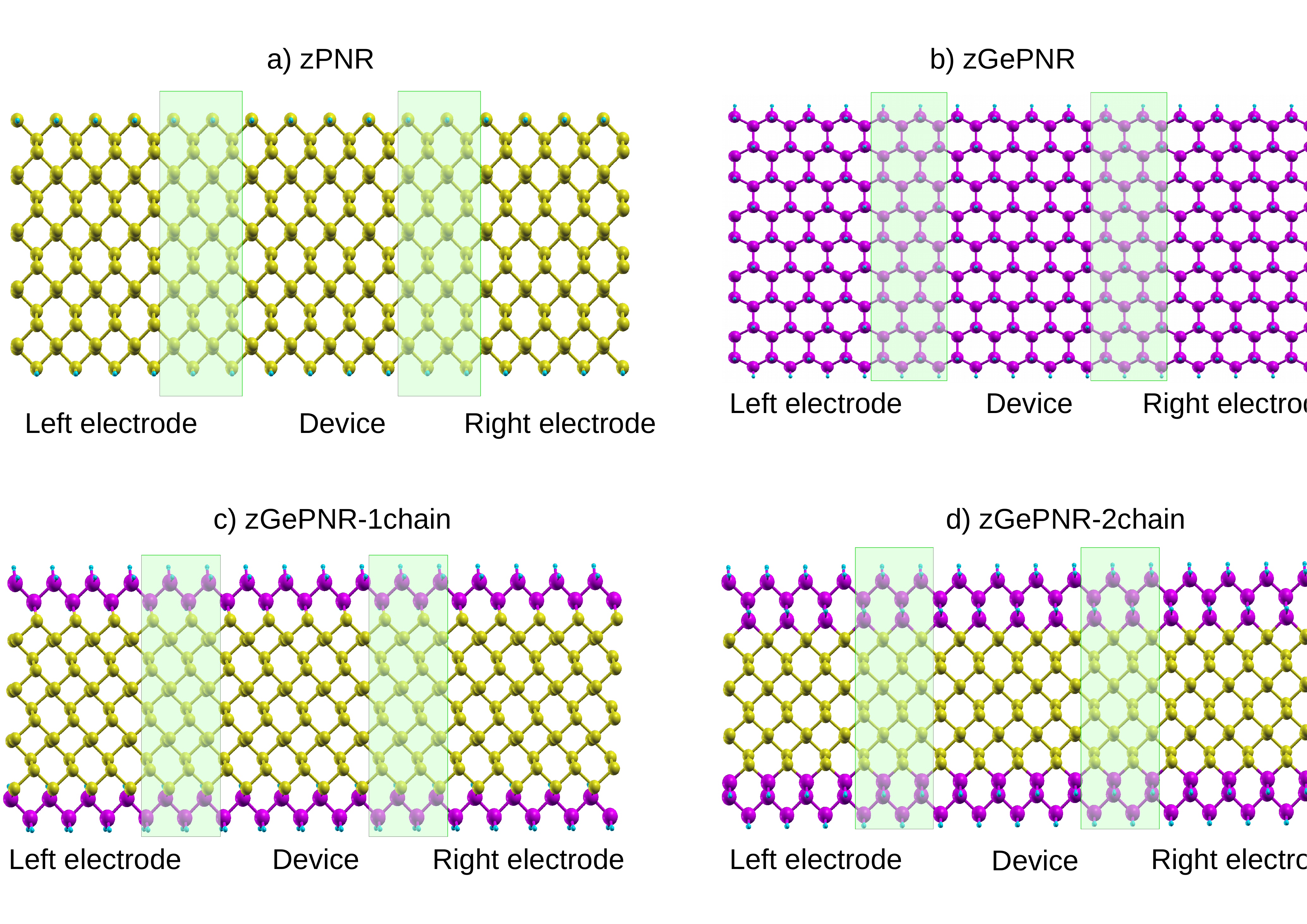}
	\caption{\label{structure}
		(Color online) The computational setup for a) zPNR, b) zGeNR, c) zGePNR-1chain and d) zGeNPR-2chain.
		phosphorus, germanium and hydrogen are in yellow, purple and blue spheres respectively.} 
\end{figure}
\end{center}
\end{widetext}

\subsection{Computational details}
The first principle calculations are performed within the
Density Functional Theory(DFT) method, as implemented
in SIESTA open-source package\cite{soler2002siesta}.\\

We assume the infinite length and priodic boundary conditions in the x-direction. 
To avoid interactions between the adjacent layer,  more than $20A^\circ$ vacuum space considered to separate the nano-ribbons in two other directions, . \\

Spin-unrestricted DFT calculations are performed using the Perdew Burke Ernzerhof (PBE) for generalized gradient approximation (GGA) exchange and correlation functional approach to obtain the band structures for the selected structures. 

In our calculations, Brillouin zone is sampled with $50 \times 1 \times 1$ k-point grid and the cutoff energy is fixed to be 150-Ry.

The double-$\zeta$ polarized basis sets are used for the valence band electrons. The structures are relaxed until the interatomic forces are less than  $0.002eV /A^\circ$. 

In order to perform the calculations of charge transport
and electrical properties of the systems, the Non-equilibrium Green's Functions (NEGF)\cite{taylor2001ab} equations are solved using the Kohn-Sham wave functions obtained from DFT,\cite{brandbyge2002density} as implemented in TRANSIESTA open-source package\cite{stokbro2003transiesta} at the room temperature.\\
To investigate the transport properties, we specify three regions within the sample: two electrodes and the central (device) region (Fig.\ref{structure}).




The transmission function of the system is calculated according to the following equation:\cite{datta1997electronic}
\begin{equation*}
T(E,V_b)=Tr[\Gamma_L(E,V_b)G(E,V_b)\Gamma_R(E,V_b)G^\dagger (E,V_b)]
\end{equation*}
where $E$, $V_b$ and $G$ are energy, bias voltage and green function, respectively. $\Gamma_{L(R)}$ is the spectral density describing the coupling between the right (left) electrode and the scattering region.

In these structures current vs bias voltage is extracted by the Landauer- B\"{u}ttiker\cite{landauer1957spatial,buttiker1986four} formula, 
\begin{equation*}
I(V_b)=\frac{2e}{h}\int_{-\infty}^{\infty}dET(E,V_b)[f(E-\mu_L)-f(E-\mu_R)]
\end{equation*}
where $\mu_{R(L)}$ is the chemical potential of the right (left) electrode, $eV_b = \mu_L-\mu_R$ and $f(E-\mu_{R(L)})$ is the Fermi function.\\
We change the bias voltage from $0.0 V$ to $2.0V$ with the step of $0.05 V$ to achieve convergence of the density matrix.

 \begin{widetext}
 	\begin{center}
 		\begin{figure}[htb]
 			\begin{center}
 				\centering
 				\includegraphics[scale=0.6,center]{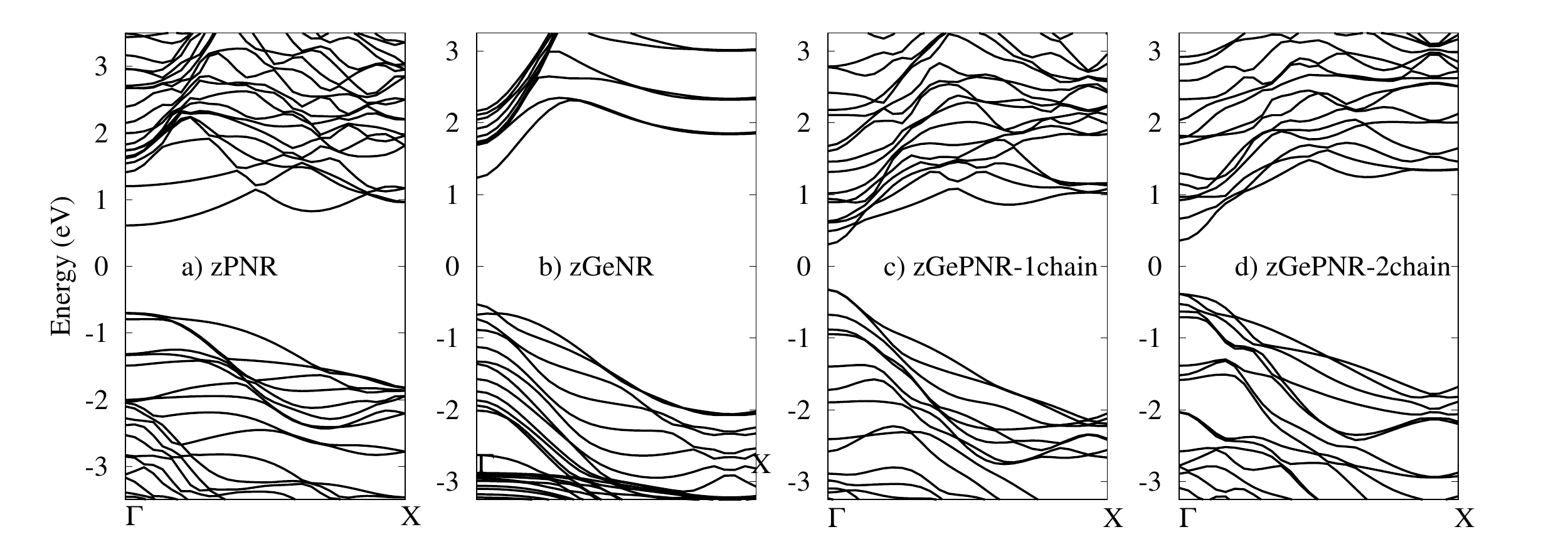}
 				\caption{\label{band}
 					(Color online) The band structures of a) zPNR, b) zGeNR, c) zGePNR-1chain and d) zGeNPR-2chain..
 				} 
 			\end{center}
 		\end{figure}
 	\end{center}
 \end{widetext}
 
\section{Results and discussion}
\label{sec:results}
The band structures of the studied configurations are illustrated in Fig.\ref{band}. 
In the case of phoshorene nanoribbon, each phosphorus atom is covalently bonded to three 
other P atomes and forms a $sp^3$ hybridization to construct a puckered honeycomb structure with a DFT bandgap about $1.3eV$. 
\begin{figure}[ht]
	\includegraphics[width=90mm]{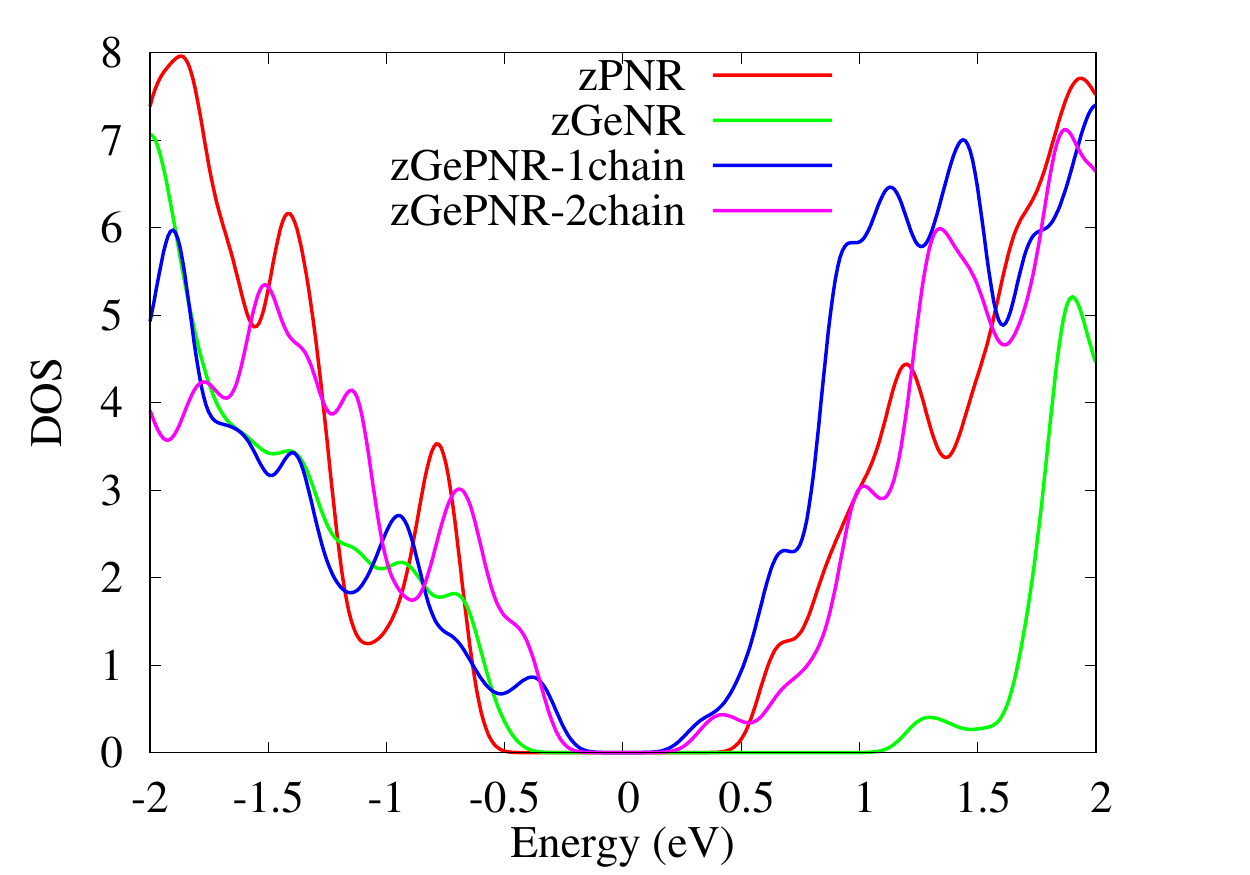}
	\caption{\label{Band_DOS}
		(Color online) Density of states for zPNR, zGeNR, zGePNR-1chain and zGeNPR-2chain.} 
\end{figure}
The chemical bonds in germanium are actually of $sp^3$ -like\cite{wang2011comparative} type for the partial hybridization of $s$ and $p_z$ in Ge but it has a larger covalent radius and its standard atomic weight is two times bigger than that for P. All this specifications along with its electronegativity which is almost of the same order as in the case of P, result in DFT bandgap of $1.67~eV$ which is not very far from P DFT bandgap. 

As shown in Fig.\ref{Band_DOS}, zPNR (zigzag edge phoshorene nanoribbon) has a higher Density of States (DOS) in the region close to Fermi energy in comparison to zGeNR (zigzag edge germanene nanoribbon). This explains why zGeNR has bigger DFT bandgap. 

This is true for the two other configurations studied in the present work. Replacing edge phosphorus atoms by germanium atoms, Fig.\ref{structure}c,  leads to increase in zGePNR-1chain's DOS around Fermi energy. As a result, the DFT bandgap is decreased in comparison to pure zPNR and zGeNR, Fig.\ref{band}. Going further and replacing the next series of phosphorus atoms by Ge ones, Fig.\ref{structure}d,  zGePNR-2chain shows less DOS around Fermi energy. This is the reason for zGePNR-2chain to have a relatively larger DFT bandgap rather than zGePNR-1chain. \\ 
%

\begin{figure}[tb]
	\includegraphics[width=90mm,center]{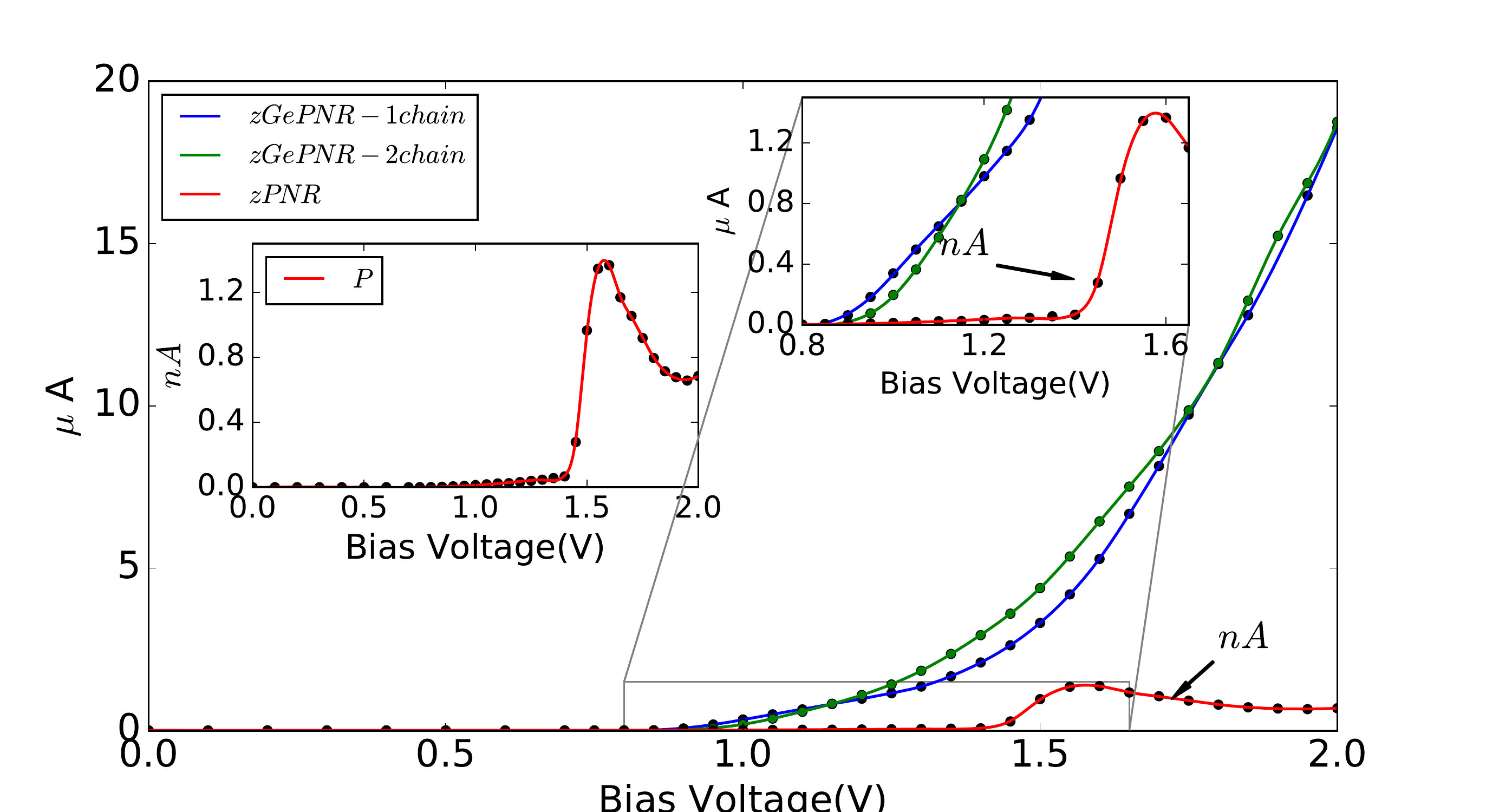}
	\caption{\label{V_I}
		(Color online) \textit {I -V} characteristics of zPNR, zGePNR-1chain and zGeNPR-2chain.} 
\end{figure}

To investigate the charge transport in our systems, we determine the current-voltage bias characteristic of the configurations. 
As shown in Fig.~\ref{V_I}, 
%
nonlinear \textit{I-V}-traces characteristic for tunneling appear at 
elevated bias. 

One could notice that for bias voltages lower than $1 V$ the current nearly vanishes 
for three systems while it increases 
with a relatively fast slope for higher bias voltages.\\
In fact, applying a bias voltage
shifts the Fermi level of the left electrode with respect to the
Fermi level of the right one. The current starts flowing
once the top of the valence band of the left electrode matches
in energy with the bottom of the conduction band of the right
electrode. 
This occurs in both systems, but for zPNR is about three orders of magnitude smaller than for germanium doped zPNR and it rapidly increases at about $1.4 V$. 

The $I-V_{bias}$ curve can be divided into three interesting regions. In the first one, up to $1.2 V$ both configurations (zGePNR-1chain and zGePNR-2chain) show the same behavior with almost the same values of currents. With increasing the bias voltage, they start to split with small differences up to $1.75 V$. Then, again they behave in almost the same way. 

\begin{figure}[tb]
	\includegraphics[width=90mm]{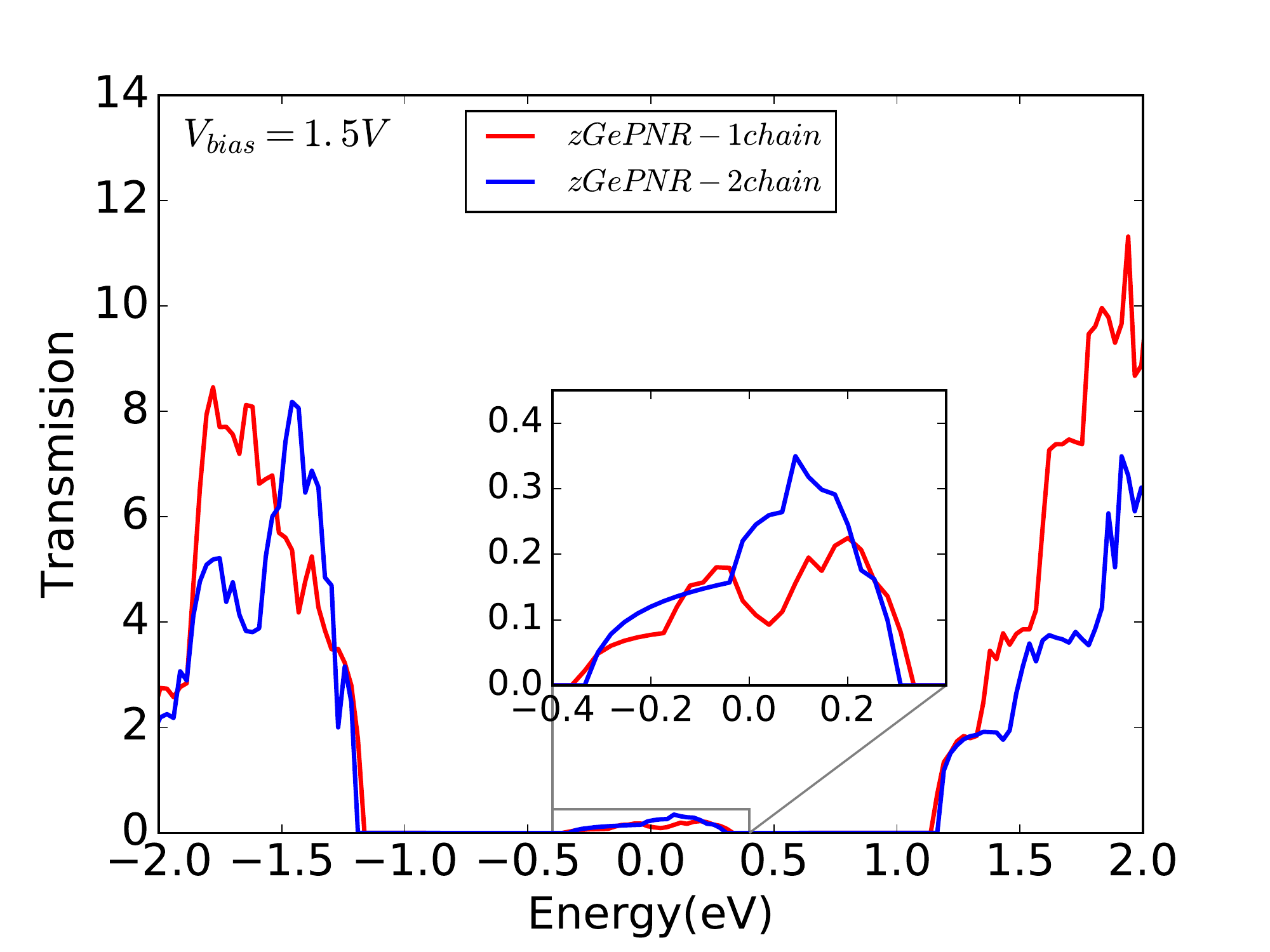}
	\caption{\label{trans}
		(Color online) Transmission spectra of zGePNR-1chain and zGePNR-1chain at $V_{bias}=1.5~V$} 
\end{figure}

This is confirmed in Fig.~\ref{trans}, where transmission spectrum of both configurations is shown at $V_{bias} = 1.5~V$. In the range of bias voltage, between $1.2~V$ and $1.75~V$ the overlap of DOS for both electrodes and scattering region in zGePNR-2chain increases. This results in more eigenchannells open and thus larger number of electrons transfered. That is why the $I-V_{bias}$ curve shows higher values of current for zGePNR-2chain in this bias voltage range. 

For the zPNR system, the current
increases with the applied bias voltage and reaches the
maximum value of $1.36~ \mu A$ at $V_{bias} = 1.55~V$. However, when $V_{bias}$ 
increases further, the current decreases dramatically, and
consequently the negative differential resistance (NDR) phenomenon arises.

An important aspect is the path through
which the current flows in the channel material. We have traced
the pathway for the current by studying the eigenvectors of the
transmission spectrum for different bias voltages.

\begin{figure}[tb]
	\includegraphics[width=90mm,center]{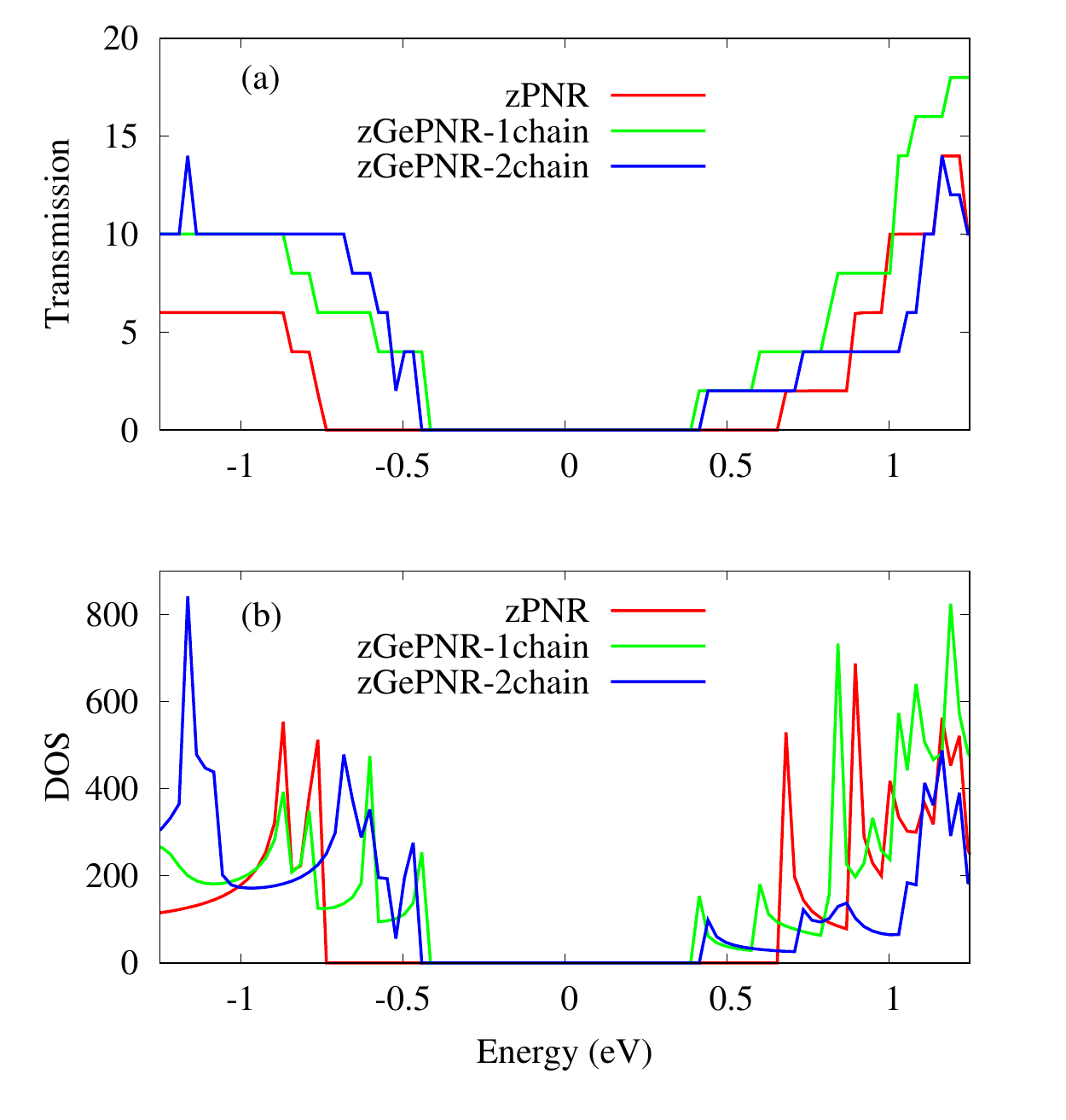}
	\caption{\label{Dos-T-0V}
		 (Color online) a) Transmission spectra and b) Density of states for the three configurations at $V_{bias}=0~V$}
\end{figure}
According to Fig.~\ref{Dos-T-0V}, the bandgaps at zero bias voltage are of about $1.3 eV$, $0.63 eV$ and $0.74 eV$ for zPNR, zGePNR-1chain and zGePNR-2chain, respectively which 
shows that the presence of germanium decrease the bandgap and causes corresponding increase in the transmission spectrum and 
charge transport, Fig.~\ref{Dos-T-0V}a.

In fact, at zero bias voltage, both electrodes have exactly the same DOS, 
and hence the transmission function is large wherever the electrodes have electronic states. 
However, since the bias window is zero, there is no current, as it is confirmed by Fig.~\ref{V_I}.\\
Fig.~\ref{Dos-T-0V}b, shows the DOS for zPNR, zGePNR-1chain and zGePNR-2chain and it can be seen 
that there exists no overlap between the electrodes in the absence of bias voltage.

By increasing $V_{bias}$, the chemical 
potential of electrodes changes and it reaches to the 
region where the overlap between the states in the electrodes 
and the device is not zero anymore. In this case, the charge 
transport takes place as it is 
shown in Fig.~\ref{Dos-TE-1-2}, for germanium doped PNR.
\begin{widetext}
 \begin{center}
\begin{figure}[tb]
	\includegraphics[scale=0.7,center]{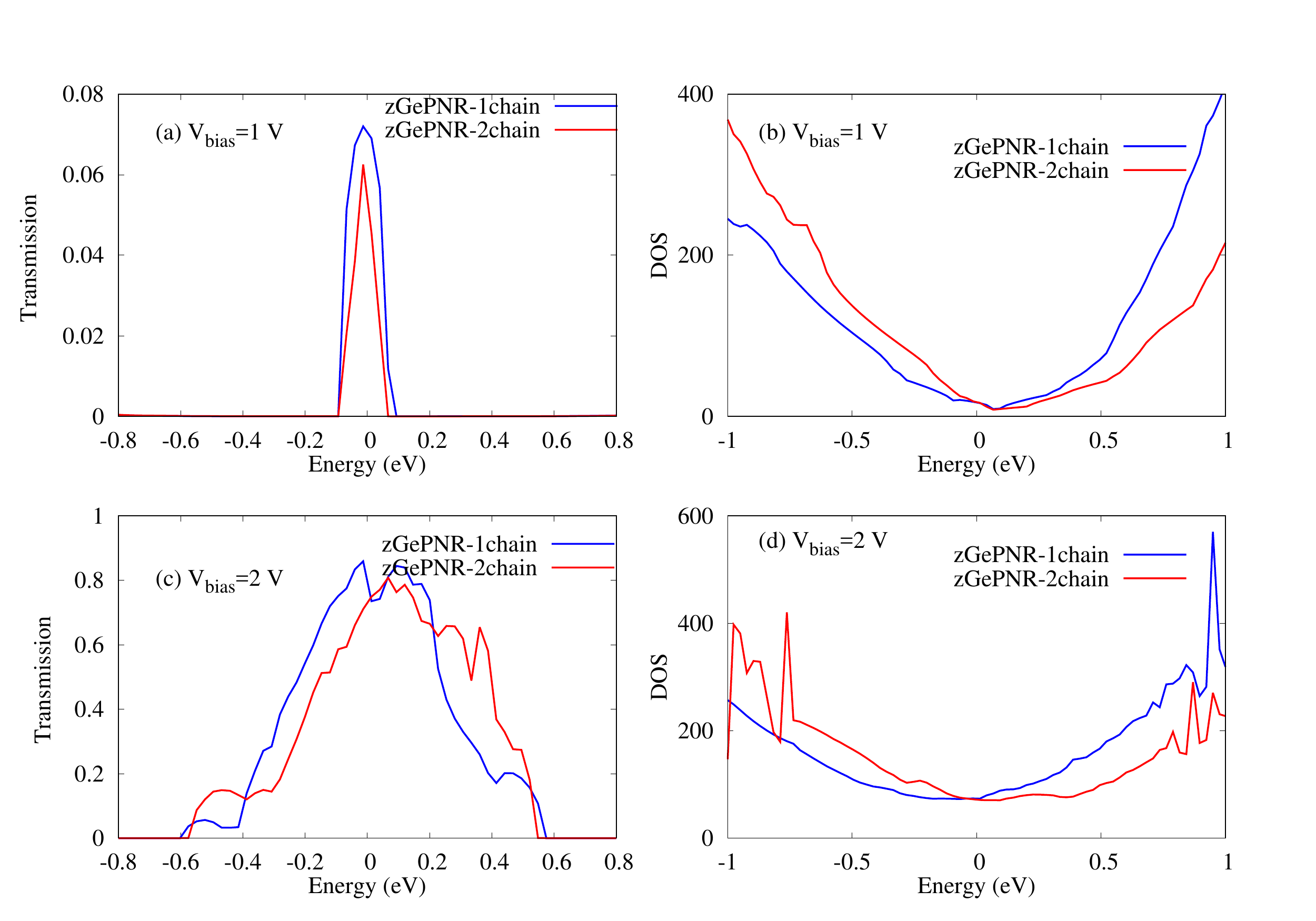}
	\caption{\label{Dos-TE-1-2}
		(Color online) Transmission and Density of States for zGePNR-1chain and zGePNR-2chain at $V_{bias}=1~V$ and $V_{bias}=2~V$ respectively.} 
\end{figure}
\end{center}
\end{widetext}
The upper panel of Fig.~\ref{Dos-TE-1-2} shows the transmission spectrum and  
the Density of States (DOS) at the Fermi energy of the device for a
bias voltage of $1 V$. One can see that the amplitude of transmission spectrum is large
only at the two edges of DOS. Thus it is clear that the current is carried
by the edge states, and there is virtually no current in the
central region.

The DOS has large amplitude near the left electrode.
The amplitude decreases as we move from the left 
electrode into the device and near the right
electrode the amplitude nearly vanishes resulting in a very small current 
(Fig.~\ref{V_I}(a) shows almost zero value for current at $V_{bias}=1~V$).  

Lower panel of Fig.~\ref{Dos-TE-1-2} represents the transmission spectrum and  
the Density of States (DOS) for a bias voltage of $2 V$. One can notice that the 
amplitude of transmission spectra in the bias voltage interval increases, Fig.~\ref{Dos-TE-1-2}c. This results 
from the overlap between DOS in both electrods and the device, Fig.~\ref{Dos-TE-1-2}d.

\begin{figure}[tb]
\includegraphics[scale=0.5,center]{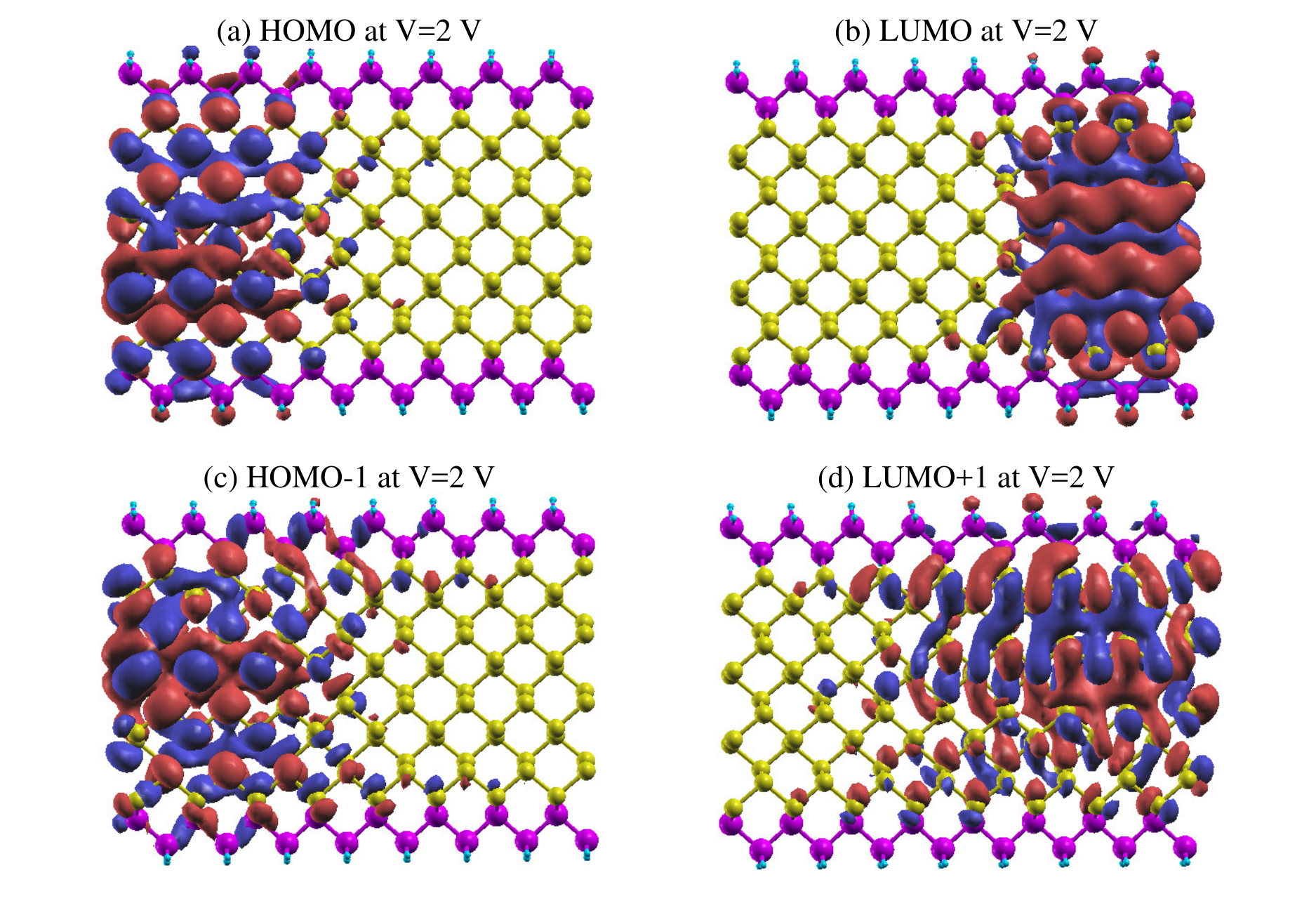}
\caption{\label{GeP_DOS_HL}
(Color online) The HOMO, LUMO, HOMO-1 and LUMO+1 MPSH orbital for zGeRNR-1chain at $V_{bias}=2 ~V$.} 
\end{figure}

To better understand the effect of $V_{bias}$, we show the eigenstates of the molecule placed in two different probe 
environments. The eigenvalues were calculated according the molecular projected self-consistent Hamiltonian (MPSH), Fig.~\ref{GeP_DOS_HL}. 
The clearly nonzero transmission spectra emerges around the 
Fermi level and the isosurfaces appear inside the device. 
Fig.~\ref{GeP_DOS_HL} shows the highest occupied molecular orbital (HOMO) 
and lowest unoccupied molecular orbital (LUMO) for the first (Fig.~\ref{GeP_DOS_HL}a and b) and second (Fig.~\ref{GeP_DOS_HL}c and d) molecular 
orbital levels for zGePNR-1chain at $V_{bias} = 2~V$. The state overlap can 
be clearly seen in the right upper panels where a non-zero current flows from 
left electrode to the right one (Fig.~\ref{V_I}(a) shows $4.4 \mu A$). The good compatibility can be seen 
between Fig.~\ref{GeP_DOS_HL}(a) and Fig.~\ref{GeP_DOS_HL}(c). 
This is what one would expect in a sizable bandgap semiconductor.
The applied bias voltage changes the eigen-channels 
in the electrode-device interfaces and the bandgap narrowing occurs due to induced electrostatic potential 
across the nanoribbon.

The calculated molecular projected self-consistent Hamiltonian (MPSH) (Fig.~\ref{GeP_DOS_HL}) shows the eigenstates of the molecule
which placed in a two-probe environment. This confirms that the spatial
distribution of orbital levels has been affected by the electrodes.

\begin{figure}[tb]
\includegraphics[scale=0.5,center]{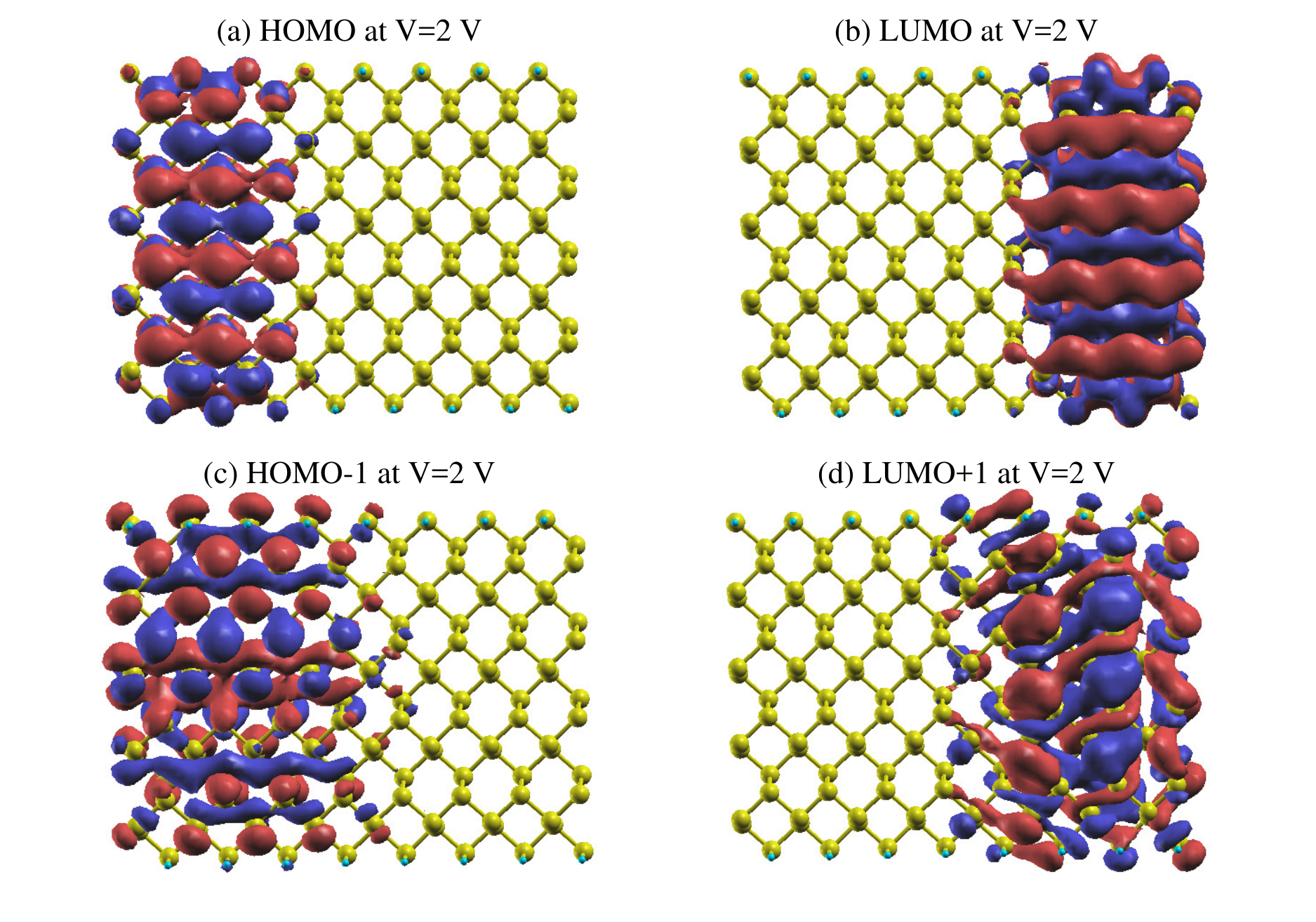}
\caption{\label{P_DOS_HL}
(Color online) The HOMO, LUMO, HOMO-1 and LUMO+1 MPSH orbital for zRNR at $V_{bias}=2 ~V$.
} 
\end{figure}

The same effect occurs for zPNR, Fig.~\ref{P_DOS_HL}. The only difference is that for zPNR the maximum amplitude of current 
is about three orders of magnetide smaller. 

This allows to explain the DOS spectra shown in Fig.~\ref{Dos-T-0V}. If Ge is embeded in the structure, 
the spectra penetrate deeper into the conduction region (about $1~eV$) than those for zPNR. 

%


In order to investigate the calculated rectification behavior, we studied the transmission spectrum $T (E, V_{bias})$ 
as a function of energy $E$ and the bias voltage $V_{bias}$, Fig.~\ref{P_GeP_DOS_Trans}a, which we compare the corresponding DOS and transmission spectra of zPNR and 
germanium doped PNR. 
As seen in Fig.~\ref{P_GeP_DOS_Trans}, transmission 
happens to some extend in both structure, however in zPNR the amplitude is almost zero even in the highest $V_{bias}$.
\begin{widetext}
 \begin{center}
\begin{figure}[H]
\includegraphics[scale=0.7,center]{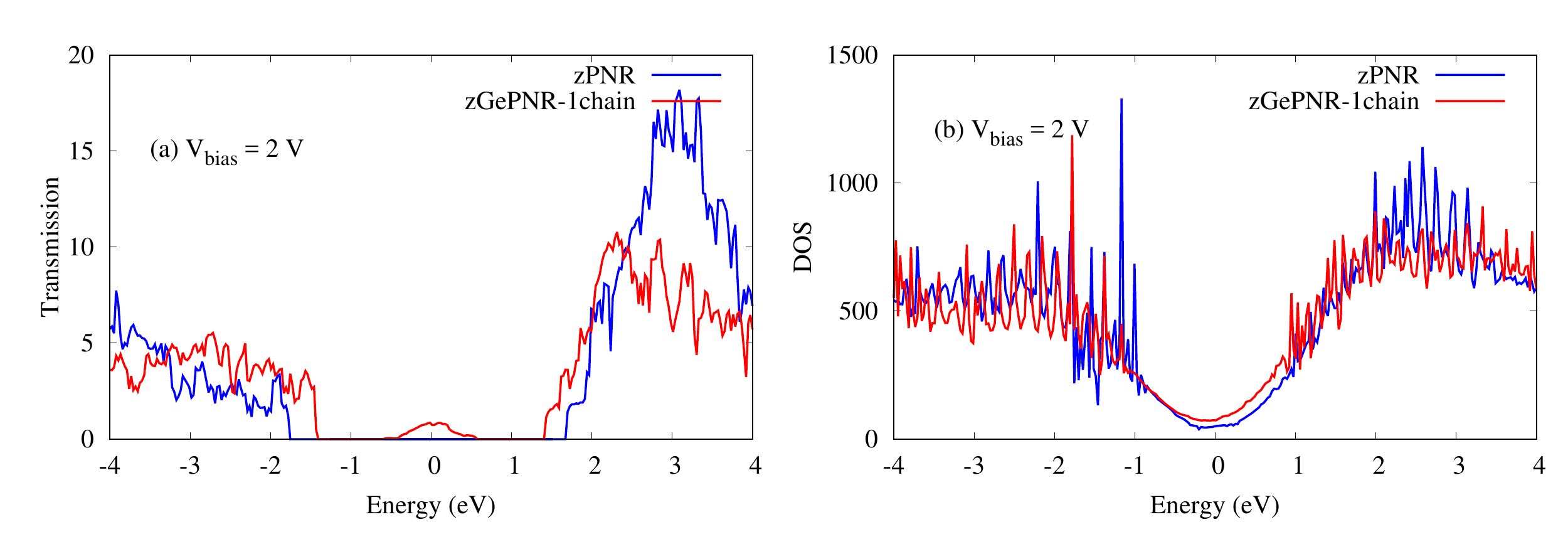}
\caption{\label{P_GeP_DOS_Trans}
(Color online) a) Transmission spectrum and b) Density of States for zPNR and zGePNR-1chain at $V_{bias}=2 ~V$.
} 
\end{figure}
\end{center}
\end{widetext}

The calculated DOS shows a peak at the Fermi level, which implies a strong correlation between
transmission and DOS. This occurs because the transport at the Fermi level is dominated by resonant
tunneling through interface states (see Fig.~\ref{P_GeP_DOS_Trans}), not barrier
tunneling\cite{Qingyun}.

Under a positive bias voltage the chemical potential of the left (right) electrode 
shifts by $\frac{eV}{2} (-\frac{eV}{2})$. Therefore, the DOS of the left (right) electrode 
shifts toward higher (lower) energy by $\frac{eV}{2}$, Fig.~\ref{P_GeP_DOS_Trans}. 



From Fig.~\ref{P_GeP_DOS_Trans}, one expects to have higher transmission for the 
higher applied bias voltage. This is confirmed by Fig.~\ref{contour}, which 
proves increasing the bias voltage causes higher transmission values (lighter color) in both structures.

The two solid lines in Fig.~\ref{contour} show the total current obtained from 
the integration of the transmission function in a given bias window. 
The positive (negative) transmission refer to electron and hole conductivities, respectively. 
One may notice that in positive (negative) region, the 
contributions of electrons (holes) in conductivity dominate. As can be seen in Fig.~\ref{contour}b and Fig.~\ref{contour}c 
replacing P edge atoms by Ge atoms, results in the increased transmission and thus also conductivity and current.

\begin{widetext}
\begin{center}
\begin{figure}[tb]
	\includegraphics[scale=1.5,center]{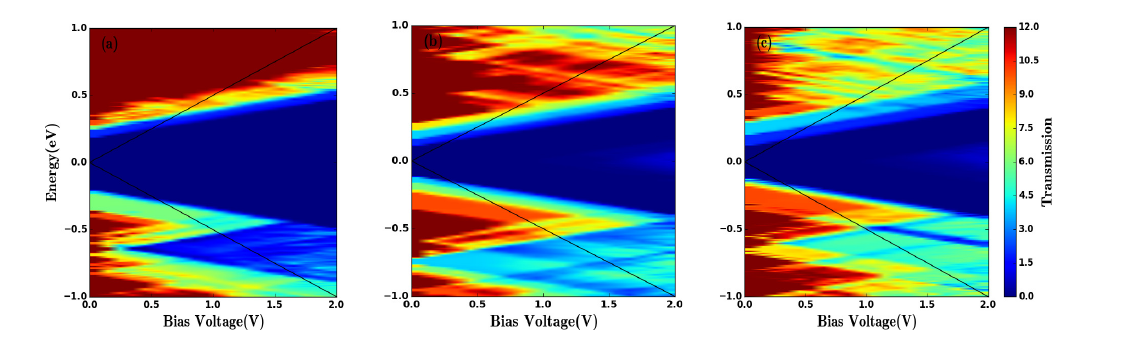}
	\caption{\label{contour}
		(Color online) The transmission spectra of a) zPNR, b) zGeNPR-1chain and c) zGePNR-2chain as a function of charge carrier 
		energy and bias voltage.} 
\end{figure}
\end{center}
 \end{widetext}
 
\section{Conclusions}
\label{sec:concl}
We studied two new systems consisting of phosphorus, germanium and hydrogen atoms in a zigzag nanoribbon. 
The structures were relaxed using DFT method under 
exchange-correlation potential GGA.

The electronic and transport properties of the systems are investigated 
within the noneqilibrium Green's function formalism and density functional theory.

DOS and $T(E,V_{bias})$ were investigated and show that the charge transport occurs 
when the bias voltage reaches to about 1 \textit{V}. 

First, the electronic structures of pure zPNR and zGeNR were analyzed then 
we showed how replacing edge phosphorus atoms with germanium atoms results in decreasing of the 
bandgap of zPNR. 

The transport channels are studied via the calculations of 
the current density and local electron transmission pathway.

The calculated MPSH shows that the spatial
distribution of orbital levels was affected by the electrodes. 
The visualized transmission pathways show how charge carriers propagate through the 
scattering region in all systems. 

The characteristics of the new structure were compared to those of zPNR. 
The results confirm that the bandgap of phosphorene nanoribbon in the 
presence of germanium decreases which results in the increase of charge transport. 

The studied structure has a bandgap of about 0.7 \textit{eV} which absorbs light in 
the visible range and thus is an interesting contender for solar cells. 
The characteristics of the present systems make them suitable for 
practical applications in nanoelectronic and optoelectronics and devices.

Finally, our calculations show that negative differential resistivity behavior in zPNR device vanishes when the 
edge phosphorus atoms are substituted by Ge ones. 

\begin{center}
\normalsize \textbf{Acknowledgments}
\end{center}
This work was partially supported by Iran Science Elites Federation grant.

\bibliographystyle{prsty}
\bibliography{refrences.bib}

\end{document}